\documentclass[preprint2]{aastex}
\usepackage{epsfig}

\shorttitle{Supegalactic winds}  
\shortauthors{} 
 
\begin{document} 

\title
{Supergalactic winds driven by multiple superstar clusters.}

\author{Guillermo Tenorio-Tagle \altaffilmark{1},
        Sergiy Silich \altaffilmark{1} and 
        Casiana Mu\~noz-Tu\~n\'on \altaffilmark{2}}
\altaffiltext{1}{Instituto Nacional de Astrof\'\i sica Optica y Electr\'onica, 
                 AP 51, 72000 Puebla, M\'exico}
\altaffiltext{2}{Instituto de Astrofisica de Canarias, E 38200 La Laguna, 
                 Tenerife, Spain}

\begin{abstract} 
Here we present two dimensional hydrodynamic calculations of free expanding 
supergalactic winds, taking into consideration strong radiative
cooling. Our main premise is that supergalactic  winds are powered by  
collections of superstar clusters, each of which is  a source
of a high metallicity supersonic diverging outflow. The interaction of 
winds from neighboring superstar clusters is here shown to lead to a 
collection of stationary oblique shocks and crossing shocks, able to 
structure the general 
outflow into a network of dense and cold, kpc long filaments that originate
near the base of the outflow. The shocks  also lead to extended
regions of diffuse soft X-ray emission and furthermore, to channel the 
outflow with a high degree of collimation into the intergalactic medium.  

\end{abstract} 
 
\keywords{Galaxies: superwinds, Starbursts: optical and X-ray               
emissions, general - starburst galaxies}

\section {Introduction} 
 
Supergalactic winds (SGWs; see Heckman 2001), as first envisaged by Chevalier and Clegg 
(1985; hereafter CC85), freely flow from the nuclear starburst region 
of a galaxy into the intergalactic space. The flow originally thought 
to behave in an adiabatic manner has recently been shown to be
strongly affected by radiative cooling, particularly for powerful 
($\geq 10^{41}$ erg s$^{-1}$) and compact starbursts (see Silich et al. 2003;
hereafter referred to as Paper I). Radiative cooling does not disturb
the velocity of the outflow, nor the expected density drop 
($\sim$ r$^{-2}$) acquired as the flow moves away from the massive 
central cluster. Rapid cooling however, brings the temperature down in regions close to the starburst, favoring 
 rapid recombination, and thus making the fast streaming 
outflow easy target of the UV radiation generated by the central
stars. The prediction is then a much reduced size of the X-ray emitting zone 
and a fast moving HII region gas originating close to the central starburst. 

M82 with its wind striking the "H$\alpha$ cap" at 10 kpc from the 
central source (see Devine \& Bally 1999), 
is without a doubt the best example of a SGW in the Local Universe, injecting
its newly processed metals into the intergalactic space (IGM). The 
central biconical wind in M82 however, presents an intricate structure
that has little to do with the outcome from models of superwinds. In 
particular the X-ray and the H$\alpha$ filamentary structure as well
as their spatial coincidence have presently no explanation. As stated by 
Strickland \& Stevens (2000) the predicted X-ray luminosity, even 
using the adiabatic model of CC85, falls short by more than three
orders of magnitude below the observed value. Strickland et al. (1997) 
also showed that the entropy of the X-ray emitting gas increases 
with distance from the plane of the galaxy, fact that is inconsistent 
with an adiabatic outflow model.  On the other hand, the H$\alpha$ 
filamentary structure, beautifully evidenced by Subaru 
(see Ohyama et al. 2002 and references therein),
is clearly not limb brightened and originates right at the base of 
the outflow reaching several kpc in a  direction almost perpendicular to
the galaxy plane. These facts are also very different from the 
superbubble features calculated by Suchkov et al. (1994) in which 
a complex filamentary structure develops at large distances 
(several kpc) from the galaxy disc, as a result of matter entraining 
the hot superbubble. There should be in fact very little resemblance 
between supergalactic winds freely expanding into the IGM and the 
calculations mentioned above in which a supershell contains always the 
possible outflow of the superbubble interior, and thus inhibits the 
development of a super galactic wind.  

So far, all calculations in the literature have assumed that the
energy deposition arises from a single central cluster  that spans 
several tens of pc, the typical size of a starburst. However,
recent optical, radio continuum, IR and UV observations (Ho 1997; 
Johnson et al. 2001; Colina et al. 2002; Larsen \& Richtler 2000, 
Kobulnicky \& Johnson 1999) have revealed a number of unusually
compact young stellar clusters.
These overwhelmingly luminous concentrations of stars present a
typical half-light radius of about 3 pc, and a mass that ranges from several 
times 10$^4$ M$_{\odot}$ to a few 10$^6$ M$_\odot$. Clearly these units
of star formation (superstar clusters, SSCs) are very different to what
was usually assumed to be a typical starburst. Several of 
these entities have now been identified within a single starburst 
nucleus. For example, there are about 100 of them composing a
flattened distribution of 150 pc radius at the center of M82, 
(see de Grijs et al. 2001, O'Connell et al. 1995), at least four
within the nuclear zone of NGC 253 (Watson et al. 1996), and many of 
them in the Antenae (Whitmore et al. 1999).

Here we investigate the effects that the presence of several of these 
young compact clusters in a galaxy nucleus 
may have on the inner structure of well developed, or freely expanding, SGWs.
Several aspects are considered in this two dimensional, first approach, to 
the 3-D interaction of multiple powerful winds. Among these, the metallicity
of the superwind matter is here shown to have a profound impact on the inner 
structure of SGWs.

Here we study the interaction of several strong winds emanating from
a collection of nearby superstar clusters, causing together the 
development of an extended region in which a plethora of crossing
shocks collimate the general outflow while giving rise to an important 
soft X-ray contribution at large distances from the starburst nuclei.  
An extended  region in which the layers of strong direct wind 
interactions lead, under strong radiative cooling, to a well developed 
network of elongated filaments.  Section 2 displays our two
dimensional calculations that use CC85 as initial condition in each 
of the superstar clusters. The evolution leads to multiple
interactions, and the outcome of these, depending on the local 
values of density, temperature and metallicity, define when the 
flow is affected by  strong radiative cooling. Section 3 discusses
some of the observational consequences of such well collimated and structured  
outflows, in particular the resultant filling factor for different gas 
phases.

\section {The interaction of winds from superstar clusters} 

There are two different evolutionary stages in the remnants produced
by the large mechanical energy input rate 
associated to nuclear starbursts. The first one, the superbubble
phase, is that during which  
a large-scale remnant evolves within the ISM, either within the disk
or within the halo of the host galaxy.
The second phase, the freely expanding supergalactic wind, in which an 
open  channel in the ISM allows 
the supernova matter to freely stream into the intergalactic medium.

Weather one assumes a central source or multiple SSCs composing a starburst nucleus, the initial
interaction with the host galaxy ISM will be rather similar. 
In particular, both sources of energy will lead to the development of 
a leading shock, able to sweep the surrounding ISM into a large-scale supershell, while a reverse shock
will cause the full thermalization of the matter violently ejected by the central source or the multiple stellar clusters 
(see for example figure 1 in Heckman et al., 1990). 

In both cases the thermalized hot gas within the superbubble will 
power the leading shock and thus promote the growth of a remnant able 
to eventually exceed the dimensions of the galaxy disk, causing via 
Rayleigh - Taylor instabilities the fragmentation and disruption of 
the supershell of swept up matter (see figure 4 in Tenorio-Tagle \& 
Bodenheimer, 1988). 

If one considers the presence of an extended halo (see Silich \& 
tenorio-Tagle, 1998, 2001), the thermalized wind energy will generate 
an even larger supershell (see for example figure 6 in Strickland \& 
Stevens, 2000), the velocity of which will continuously decay as more 
halo matter is incorporated. The presence of extended haloes (see
e.g. Melo et al., 2002) is also supported by the existence of 
large-scale supershells in a large variety of galaxies (see for 
example Oey et al. 2002 and Marlowe et al., 1995).

Note that a supergalactic wind does not develop 
until the remnant exceeds the dimensions of the galaxy disk, or in case of an extended halo,
until the superbubble reaches the outskirts of the galaxy (for a review see Heckman, 2001). 
At this moment, the leading and the reverse shocks cease to exist and
the metals ejected by the star clusters will freely move into the intergalactic medium causing its
contamination.
It is also at this stage, once a channel has been carved in the ISM, once the superwind has developed, that major differences will 
arise from the assumption of a single or a multiple source of energy. 

\begin{itemize}

\item In the case of a single compact source of energy, the free expanding wind has recently been found to be
subjected to strong radiative cooling (see Paper I). Thus, the adiabatic model of Chevalier \& Clegg (1985),
used to calculate the extended X-ray emission of superwinds, leads to large overestimates.
More realistic calculations, accounting for radiative cooling, have shown a much reduced volume, of only a few times the size
of the starburst nucleus, able to generate X-rays.  In this framework it is hard to understand the extended
(more than 3 kpc long) X-ray emission of M82, a source that presents a more than 10 kpc long open channel 
into the intergalactic medium, and that clearly is in the supergalactic wind stage. 

\item Models with a single source predict also a strong H$_\alpha$ emission arising from the lateral 
walls of the disrupted supershell, continuously impacted by the UV radiation arising from the central source (see for example 
Suchkov et al. 1994,  Tenorio-Tagle \& Mu\~noz Tu\~n\'on 1997, 1998). 
The emission in such a case should be strongly limb brightened. 

\item Most models (perhaps with the only exception
of Tenorio-Tagle \& Mu\~noz Tu\~n\'on 1997, 1998, which accounts for the infall of matter into the central starburst)
also end up with a remnant that presents a wide open waist
along the galaxy plane (see figures in Tomisaka \& Ikeuchi, 1988; Suchkov etal., 1994). 
This could measure several kpc in radius and is very different to the 150 pc radius estimated for the central
perturbed area of M82. 

\end{itemize}

The H$_\alpha$ emission of M82 is not limb brightened and arises from distinct filaments that emanate from the
base of the outflow. This last point is also relevant as many models attempt to explain the filamentary structure
with shell instabilities or with matter entraining the supershell at high distances from the galaxy plane (see 
Shuchkov et al. 1994).

In the following sections we derive  the properties of supergalactic winds powered by multiple superstar clusters and stress the main 
differences in the resultant structure with respect to models that assume a single source. 

\subsection{The high metallicity  outflow of supergalactic winds}

Weather one considers one or multiple sources, the amount of metals expected from massive starbursts and thus to be 
found within SGWs, depends strongly on the assumed stellar evolution models. 
 Two extreme possibilities were
investigated by Silich et al. (2001) taking into consideration models 
with and without stellar winds by Maeder (1992), Woosley et al. (1993) 
and Thielemann et al. (1993) (see also Pilyugin \& Edmunds 1996). 
The results indicate, first of all, that almost 40$\%$ of the mass gone 
into stars is violently returned to the surrounding medium by means 
of winds and supernovae. Of this, about 4$\%$ is in oxygen in the case 
of models without winds, and 1$\%$ for models that include winds. In 
either case, if one assumes that the metals mix efficiently 
with the stellar hydrogen envelopes of the progenitors, within the radius that
encompasses the stellar cluster, then the metallicity of the superwind 
can be calculated. As shown by Silich et al. (2001) for the case of superbubbles, this reaches supersolar  
values in all considered cases, showing a 
maximum within the first 7 - 8 Myr of the evolution. Such values have recently 
been confirmed through X-ray observations by Martin et al. (2002) for NGC 1569.

The large metallicities expected from massive star clusters are to have a strong impact on the cooling
properties of the outflow and thus also on the observational properties of freely expanding superwinds.   
Figure 1 shows the run of the expected 
metallicity of the matter ejected by a massive burst of stellar formation, using oxygen as tracer and stellar evolution models with winds, 
as a function of time. The estimate is for a coeval starburst 
powered during 40 Myr of evolution (until the last 8 M$_\odot$ star
explodes as supernova in a coeval starburst model). 
The outflow, before the supernova era, will display a metallicity similar 
to that of the gas cloud out of which the starburst formed. Thus the gas ejected first presents the metallicity here assumed
for the host galaxy ($Z_{ISM}$ = 0.1 $Z_\odot$). The supernova products however,
rapidly enhance the  metallicity of the outflow, reaching values well above $Z_\odot$ for at least
20 Myr of the evolution. Afterwards, once the yield is reached, the outflow steadily
approaches the metallicity values of the host galaxy. 
Starburst models, with single or multiple energy sources, are to
account for the high metallicities emanating from the massive centers 
of star formation, to derive in a consistent manner the impact
of radiative cooling in the resultant outflows.

\subsection{Boundary and initial conditions}

Several two dimensional calculations using as initial condition CC85 
adiabatic flows have been performed with the explicit Eulerian finite difference code described 
by Tenorio-Tagle \& Mu\~n\'oz-Tu\~n\'on (1997, 1998). This has been
adapted to allow for the continuous injection of multiple winds (see below).

Here we consider the winds from several identical SSCs, each with a mechanical energy 
deposition rate equal to 10$^{41}$ erg s$^{-1}$. The energy is dumped 
at every time step within the central 5 pc of each of the sources following the adiabatic solution of Chevalier \& Clegg (1985).
The separation between the sources is clearly arbitrary. We have considered two different configurations. The first 
one with three SSCs
placed at 60 pc and 90 pc from the central one, all of them 
sitting on a plane. A second configuration considers only two sources
sitting at 30 pc and 60 pc away from the symmetry axis. 
Several calculations were made to reasure that the spatial resolution used, led to  a convergent  solution.
All calculations here presented were made with the same numerical resolution of half a pc and
all of them with an open boundary along the grid outer edge.
The time dependent calculations  do not consider thermal conductivity but do account for radiative 
cooling, with a cooling law (Raymond et al. 1976) scaled to the  metallicity assumed  for every case. 
Here we present the results of different cases for which the assumed 
metallicity of the winds was set equal to 3Z$_\odot$ and 10Z$_\odot$, 
justified by the high metallicity outflows expected from massive bursts of star formation (see Figure 1).

In all cases it is assumed that at the heart of each SSC, within the region 
that encompasses each of the recently formed stellar clusters ($R_{SB}$),
the matter ejected by strong stellar winds and supernova explosions is 
fully thermalized (CC85, see also Canto et al. 2000 and Raga et al. 2001). This generates the large overpressure
responsible for the mechanical luminosity associated to each of the 
super clusters.  Within each star cluster region, the mean total mechanical 
energy $L_{SB}$ and mass ${\dot M}_{SB}$ deposition rates, 
control, together with the actual size of the star forming region 
$R_{SB}$, the properties of the resultant outflow. The total mass 
and energy deposition rates define the central temperature T$_{SB}$
and thus the sound speed $c_{SB}$ at the cluster boundary. 
As shown by CC85 at the boundary $r = R_{SB}$, the flow starts 
expanding with its own sound speed. There is however a rapid evolution 
and as matter streams away it is immediately 
accelerated by the steep pressure gradients and rapidly reaches 
its terminal velocity (V$_{\infty} \sim 2 c_{SB}$). This is due to a fast 
conversion of thermal energy into kinetic energy of the resultant 
winds. In this way, as the winds expand, their density, temperature and thermal pressure will
drop as $r^{-2}$, $r^{-4/3}$ and  $r^{-10/3}$, respectively (see CC85).
The flow is then exposed to suffer multiple interactions with 
neighboring winds and as shown in Paper I, it is also exposed to 
radiative cooling. For the former, the issue is the separation 
between neighboring sources and for the latter the local values of 
density, temperature and metallicity. Radiative cooling would
preferably impact the more powerful and more compact sources, 
leading to cold (T $\sim$ 10$^4$ K) highly supersonic streams (see Paper I).

\subsection{The structure of supergalactic winds}

Figure 2 compares the initial stages of cases 1 and 2, that consider three 
equally powerful ($L_{SB}$ = 10$^{41}$ erg s$^{-1}$) superstar clusters
sitting at 0, 60 and 90 pc from the symmetry axis. All of them with an 
$R_{SB}$ = 5 pc, produce almost immediately a stream  with a 
terminal velocity equal to 1000 km s$^{-1}$. The only difference
between the two cases is the assumed metallicity
set equal to 3Z$_\odot$ in case 1 (upper panels) and 10Z$_\odot$ in case 2 (lower panels). At t = 0 yr the 
three clusters are embedded in a uniform low density 
($\rho$ = 10$^{-26}$ g cm$^{-3}$) medium. Thus our calculations do 
not address the development of a superbubble, nor the phenomenon of 
breakout from a galaxy disk or the halo, into the IGM. The initial condition 
assumes that prior events have evacuated the region surrounding the 
superstar clusters, and we center our attention on the interaction of the supersonic outflows.

Figure 2 shows the resultant initial stages of cases 1 and 2. The various panels display 
the run of density and velocity (left panels) and that of
temperature (next four panels) in four temperature ranges: The regime of H recombination
10$^4$ K - 10$^5$ K, followed by two regimes of soft X-ray emission 
10$^5$ K - 10$^6$ K, and 10$^6$ K - 10$^7$ K and the hard X-ray 
emitting gas with temperatures between 10$^7$ K - 10$^8$ K. 

The crowding of the isocontours in the figures indicate 
steep gradient both in density or in 
temperature and velocity, and thus indicate the presence of shocks and of rapid 
cooling zones. In the temperature plots one can determine  
the distance ($\sim$ 30 pc in case 1 and 15 pc in case 2) 
 within the diverging outflows emanating from each of the superstar clusters, 
at which strong radiative cooling (in agreement with our analytical estimates in Paper I) becomes important 
in the two cases.
 
The interaction of neighboring supersonic winds causes the immediate 
formation of their respective reverse shocks,
and of a high pressure region right behind them. The pressure (and 
temperature) reaches its largest values 
at the base of the interaction plane, exactly where the reverse 
shocks are perpendicular to the incoming streams.
The high pressure gas then streams into lower pressure regions, 
defining together with radiative cooling, how 
broad or narrow the high pressure zones, behind the reverse shocks, are going to be.  
Radiative  cooling occurs in every parcel of gas at the rate 
prescribed by its local density, temperature and metallicity. If
cooling  is avoided at least partially or temporarily, 
as in the first case, the high pressure region between the reverse 
shocks would drive them against the incoming streams, and very shortly 
they would acquire a oblique standing stable configuration to be retained for 
as long as the winds continue to interact (see Figure 2, upper panels).

This also happens if cooling is fast enough,  the oblique reverse shocks
rapidly acquire a standing location, however in these cases, 
the loss of temperature behind the shocks is compensated by gas
condensation, leading, as in the second case (see Figure 2, bottom panels), 
to narrow, dense and cold filaments. The drastic drop in temperature 
 occurs near the base of the outflow, where  
the gas density is large and radiative cooling is exacerbated. The dense structures are then  launched at considerable 
speeds ($\sim$ several hundreds of km s$^{-1}$) from zones near the plane
of the galaxy (see Figure 2 lower panels).
These dense and cold structures are easy target to the UV radiation produced 
by the superstar clusters and thus upon cooling and recombination are likely to become photoionized.
Note however that as the free winds continue to strike upon these structures, 
even at large distances from their origin, the resultant cold filaments
give the appearance of being enveloped by soft X-ray emitting streams.  

All of these shocks are largely oblique to the incoming streams and 
thus lead to two major effects: a) partial thermalization and 
b) collimation of the outflow. These effects result from the fact 
that only the component of the original isotropic outflow velocity perpendicular 
to the shocks is thermalized, while the  parallel component 
is fully transmitted and thus causes the deflection of the outflow 
towards the shocks. This leads both, to an efficient  collimation
of the outflow in a general direction perpendicular to 
the plane of the galaxy, and to a substantial soft X-ray emission 
associated with the dense filamentary structure, extending up to 
large distances (kpc) from the plane of the galaxy.
In the figures one can clearly appreciate that the oblique shocks, confronting the originally diverging flows,
lead to distinct regions where the gas acquires very different temperatures, and thus that will radiate in different energy bands.  

Figures 3 and 4 show the time sequence of cases 1 (with Z = 3 Z$_\odot$) and 2 (with Z = 10 Z$_\odot$), respectively,
until they 
reach dimensions of one kpc, together with the final temperature 
structure splitted into the four temperature regimes considered earlier (last four panels). 

The stream of gas behind the reverse shocks leads eventually to 
the establishment of crossing shocks at the tips of the oblique 
structures (see Figure 3), which are to become also stationary as 
the flow is effectively channeled into the IGM. 

As in the case of colliding stellar winds from binary systems (see 
Stevens, Blondin \& Pollock 1992) a variety of dynamical instabilities 
are found to dominate the shocked region, particularly when strong 
radiative cooling sets in (see Figure 4). These lead to the various 
loops and twists along the dense filamentary structures, which 
nevertheless do not impede that the outflow reaches large distances 
away from the galaxy plane. The loops and twists along these structures 
promote also a larger cross-section to the incoming free wind and
partly thermalized wind and 
thus lead to regions of enhanced soft X-ray emission
clearly associated to the twisted H$\alpha$ filaments (see Figure 4).

Figure 5 displays the results from a final case in which two superstar 
clusters sitting at 30 and 60 pc away from the symmetry axis 
interact to shape the inner structure of a superwind. The calculation 
also assumes a Z = 10 Z$_\odot$. As in case 2, elongated filaments
result from the interaction of the high metallicity winds, channeling 
into the IGM most of the energy deposited by the SSCs. Note that, as
in the preceding cases, about 50$\%$ or less of the energy deposited 
by the most outer SSC is lost in the radial direction, while the rest, 
as well as that deposited by other energy sources, is fully
driven into the IGM.

\subsection{Self-collimated supergalactic winds}

The high degree of collimation attained in our calculations, that
composes a SGW from a collection of energetic neighboring superstar clusters,  
results from the simple fact of having placed the individual wind 
sources of equal strength, all of them, in a preferential plane. 
In this way, it becomes irrelevant if they all sit on a flattened
disk, or a ring. As long as they all sit near a preferential plane, 
the interaction of neighboring supersonic diverging flows will promote 
the multiple standing reverse oblique shocks 
and crossing shocks that will unavoidably lead to a remarkably 
efficient self-collimation. Collimation that does not required of a 
torus or a thick disk of ISM. If all superstar clusters sit on a
plane, only a fraction of 
the energy provided by the ones sitting at the most outer extremes of 
the cluster distribution will interact with the general ISM. However, 
most of the energy produced by the 
collection of sources will be rechanneled by the standing oblique shocks 
resulting from neighboring interactions,
to compose a broad base supersonic jet, capable of self-collimation.  
The base of the outflow will then have dimensions similar to the 
flattened cluster distribution and as shown above, depending on the individual
energetics, proximity and metallicity, the general outflow is to 
generate a dense and cold filamentary structure as well as a kpc extended soft
X-ray emitting region. A rich structure that could not arise if one 
assumes a free expanding wind that emanates from a single SSC. 

A wider jet structure may be generated in cases in which the 
population of SSCs do not have the same mass and thus equal mechanical 
energy input rates. Under such conditions the oblique shocks will 
present standing configurations more inclined over the less energetic 
clusters and this will lead to the  fanning and broadening 
of the outflow, and to the inclination of the filamentary pattern.

\section{Discussion}

From the starburst synthesis models (see Leitherer \& Heckman 1995) one knows 
that a  superstar cluster  with a total mass in stars 
(say $M_*$ $\sim$ 10$^6$ M$_\odot$) produces an almost constant
ionizing photon flux ($F_{uv} \sim 10^{53} $ photons s$^{-1}$) during the 
first few (3-4) Myr and then abruptly,  it
begins to decrease as t$^{-5}$  as the most massive stars begin to 
evolve away from the main sequence to eventually end up as supernovae.
This implies that after 10 Myr of evolution, the 
ionizing flux would be  more than two orders of magnitude smaller than 
its original value and thus the UV radiation will be unable to ionized 
the original HII region volume, limiting to 10 Myr the duration of the  
HII region phase. On the other hand, the mechanical 
energy deposition from such a cluster leads to an almost constant 
value ($\sim$ 10$^{40}$ erg s$^{-1}$) 
over a much longer time span, as it includes 
the correlated  supernovae from stars down to 8 M$_\odot$ with an 
evolutionary time of 40 - 50 Myr.
And thus the supernova phase is 4 or 5 times longer than the HII region phase.

Under such circumstances, if one considers a starburst nucleus composed by several SSCs generated at different times,
then the time span during which the isotropic winds from these may interact, the coherence phase, is limited to 40 Myr.  
Within this time, newly born clusters will have the capability of 
causing the ionization of the structure produce by interacting winds 
that emanate even from cluster with an age in excess of 10 Myr. 
During the coherence phase, some of the SSCs will also be
producing highly metallic outflows (see Figure 1), the interaction of 
which will lead to a filamentary wind structure.

A comparison of the last calculated time of cases 1 and 2, each with 3 
SSCs dumping 10$^{41}$ erg s$^{-1}$ (see Figures 3 and 4), when the 
redirected outflow has reached dimensions of almost 1 kpc, 
allows for an estimate of the filling factor occupied by gas at
different temperatures. The hot (T $\geq$ 10$^5$ K) gas occupies
almost 70$\%$ in case 1 and 30$\%$ in case 2, of the total area. The warm gas 
(T $\leq$ 10$^5$ K) fills most of the remaining volume although the 
dense and cold enhancements evident 
in Figure 4 occupy about 40$\%$ of the superwind cross-sectional view.
There is also a small volume around the SSCs that present 
temperatures that will allow the gas  to radiate in the hard X-ray 
regime (see last two plots in Figures 3 and 4). 
This numbers are to be compared with the results from the outflow 
produced by an equally energetic (3 $\times$ 10$^{41}$
erg s$^{-1}$) single cluster. Following  Paper I, we have calculated 
the temperature distribution and thus the radius at which 
radiative cooling (assuming Z = 3Z$_\odot$) sets in for a clusters  
with a radius of 95 pc (the size of the cluster distribution used in 
cases 1 and 2). The temperature of such a free streaming outflow
plummets to 10$^4$ K at 480 pc (instead of $\sim$ 10 kpc obtained if one 
assumes an adiabatic flow). In such a case, if one considers a similar 
volume to that displayed in Figures 3 and 4, then the hot gas 
(T $\geq$ 10$^5$ K) filling factor will be 37.5$\%$ 
and the rest of the volume will be filled with gas capable of being 
re-ionized by the stellar photon flux. When comparing the results 
from single and multiple sources, it is central to notice the spatial 
distribution of the various resultant gaseous 
phases. Cases with multiple SSCs 
lead to the spatial co-existence of the X-ray (T $\geq$ 10$^5$) 
and the dense and cold (T $\leq$ 10$^5$ K) emitting filaments 
(see panels 5 and 6 in Figures 3 and 4). In the case of a single 
source of energy the X-ray emitting gas is not in 
direct contact with the cooler medium along the outflow. i. e. the 
structure of the outflow is concentric, with the X-rays emanating only 
from the most central regions.

The collimation caused by the various oblique and crossing shocks in 
the multiple source cases, that makes the superwind avoid the
diverging outflow inherent to isotropic single source cases, 
channels almost five times more energy within the computational area 
above considered. In case 1 the thermal and kinetic energy of the hot 
phase dominate with 26$\%$ and 57$\%$, respectively,
over the 17$\%$ kinetic energy found in the gas with T $\leq$ 10$^5$ K.
These numbers are to be compared with the results from the single
source case described above that present within a similar
computational volume, a 68$\%$ and 27$\%$ as thermal and kinetic 
energy of the hot gas, while only a 5$\%$ of the total appears as 
kinetic energy of the gas with T $\leq$ 10$^5$ K.

The origin of the X-rays in nuclear starburst regions and of the 
filamentary structure, as seen in H$\alpha$ in  M82, have been
ascribed to features seen in models that consider various stages in the
development of hot superbubbles powered by a central starburst. 
Models that present a single reverse shock, an outer super shell and
thus models that have little to do with a free expanding SGW. Note that if the 
wind of M82 has reached the H$\alpha$ cap at 10 kpc from the
galaxy disk, and is expanding with say, 1000 km s$^{-1}$,
then the free streaming outflow started at least 10 Myr ago. During 
that time the base of the outflow has managed to preserve a
comparatively small dimension (radius $\sim$ 150 pc) implying a very 
efficient channeling of the deposited energy in a direction away from 
the disk of the galaxy.  
There is also a growing evidence of large-scale features in starburst galaxies,
caused by an important stellar energy input rate and the richness of 
structure within the ISM. However, in most of the cases the evidence 
for a freely expanding supergalactic wind is only marginal. In the
case of NGC 253, its extended X-ray emitting bubble is much smaller 
than the dimensions of the dusty galaxy halo found by Melo et
al. 2002, implying therefore that it is still evolving along the 
superbubble phase. Similar conclusions were drawn by Martin et al. 
(2002) for NGC 1569: "The X-ray color variations 
in the halo are inconsistent with a free-streaming wind and probably 
reveal the location of shocks created by the interaction of the wind 
with a gaseous halo". Several more dwarf starburst galaxies were 
considered by Legrand et al. (2001) where the energetics 
inferred from the central clusters were compared with the limit for 
mass ejection derived by Silich \& Tenorio-Tagle (2001). In all  
considered cases the bulk of the galaxy sample lie below the limit 
required to reach the galaxies outskirts.

From the results of section 2, it is clear that the inner structure of super
galactic winds strongly depends on the energy and mass deposition
history. In particular we have shown that the richness of structure is 
largely enhanced when the presence of superstar clusters, their
powerful winds, and possible interaction within a single starburst 
nucleus, are taken into consideration. These considerations open a 
new set of possibilities. Issues such as the intensity of star
formation in every superstar cluster, which defines their mechanical 
luminosity, their age which also impacts on the metallicity of the 
ejected matter,  as well as the number of superstar clusters, their 
compactness, and their position within a starburst nucleus, are all 
relevant new parameters that allow for the co-existence of X-rays and 
optical emitting features, even at large distances from the source of energy.

From our results it is clear that a plethora of structure, both in 
X-rays and in the optical line regime, may originate from the 
hydrodynamical interaction  of multiple winds. The interaction leads 
to multiple standing oblique (reverse) shocks and crossing shocks 
able to collimate the outflow away from the plane of the galaxy. 
In our two dimensional simulations, these are surfaces that 
become oblique to the incoming streams and thus evolve into oblique 
shocks that thermalize only partly the kinetic energy
of the winds causing a substantial 
X-ray emission at large distances away from the galaxy plane. 
Surfaces that at the same time act as collimators, redirecting the winds 
in a direction perpendicular to the plane occupied by the 
collection of SSCs. Radiative cooling behind the oblique shocks 
leads, as soon as it sets in, to condensation of the shocked gas, 
and thus to the natural development of a network of filaments that 
forms near the base of the outflow, and stream  away from the plane 
of the galaxy to reach  kpc scales. Under many circumstances these 
filaments develop right at the base of the outflow and for all cases the 
prediction is that they are highly metallic. Note that the speed with
which the calculated filaments raise above the galaxy plane is 
$\sim 600$ km s$^{-1}$, a value in excellent agreement with the 
measured deprojected velocities of the filaments in M82 (Shopbell
\& Bland-Hawthorn, 1998). Hydrodynamic 
instabilities play also a major role on the filamentary structure.
Non-linear thin shell instabilities as studied by Vishniac (1994) 
as well as Kelvin Helmholtz instabilities broaden, twist and generally 
shape the filaments as these stream upwards and reach kpc scales. 
The broadening of the filaments causes their interaction with the 
free winds, thermalizing further the rapid stream while leading to the 
development of soft X-ray emitting zones that envelope the densest structures.

The surfaces that develop at the plane of interaction between two wind 
sources are in our two dimensional approach depicted as vertical 
structures. From these, the only real vertical structure is the 
filament that forms along the symmetry axis in our last case. This  
results from  the convergence  of multiple winds arising from 
superstar clusters sitting on a ring 30 pc away from the symmetry axis.
It is indeed necessary to perform our calculations in three dimensions 
to see the final outcome. Note however that the lateral side of all 
interaction planes will be launched into the highest pressure regions
i.e. close to the base of the interaction, and thus it is very likely 
that they would be destroyed by the collision with other similar 
structures arising from other interaction planes. The final outcome 
is thus expected to be very similar to that depicted by our two 
dimensional simulations. Nevertheless three dimensional 
simulations are now underway and will also consider  a variety of 
stellar masses and ages of the superstar clusters, as well as 
different locations and numbers within a starburst region.

\vfil\eject

Figure captions.

\figcaption[gtt1.gif] 
{The metallicity of the ejected matter. The metallicity of freely 
streaming outflows produced by massive burst of star formation (in solar units)
is plotted against the evolutionary time.
The outflow emanates from a coeval starburst able to thermalize and mix all 
the newly processed metals with the stellar hydrogen envelopes of the 
progenitors. The curve is derived for stellar evolution models with  winds 
(see Silich et al. 2001) using oxygen as tracer. The metallicity of
the outflow reaches super solar values during most of the evolution and is 
strongly reduced down to the original assumed ISM values, once the 
production of oxygen reaches its yield.
\label{fig1}} 

\figcaption[gtt2.gif] 
{Two dimensional superwinds. The various panels in every row represent  
cross-sectional cuts along the computational grid showing: 
isodensity contours with a separation $\Delta$ log $\rho$ = 0.1 and 
the velocity field for which the longest arrow represents 10$^3$ km s$^{-1}$.
The following four panels display isotemperature contours, within the 
range 10$^4$ K - 10$^5$ K, 10$^5$ K - 10$^6$ K,
10$^6$ K - 10$^7$ K and 10$^7$ K - 10$^8$ K, respectively.
Each of the superwinds has a  power of $10^{41}$ erg s$^{-1}$ and 
a radius of 5 pc. The evolution of each wind starts from the adiabatic
solution of CC85. The distance between consecutive tick mark = 25 pc 
in all figures. Thus, the size of the plots is 100 pc $\times$ 250 pc.
The evolution time for the two models is: 1.62 $\times$ 10$^5$ yr and 
1.79 $\times$ 10$^5$ yr. The assumed metallicities were 
Z = 3 Z$_\odot$ (upper panels), and 10 Z$_\odot$ (lower panels).
\label{fig2}}

\figcaption[gtt3.gif] 
{The same as Figure 2 for Z = 3Z$_{\odot}$. The evolutionary time in  
the four first panels is 
1.62 $\times$ 10$^5$ yr, 4.17 $\times$ 10$^5$ yr, 9.4 $\times$ 10$^5$ yr
and 1.25 $\times$ 10$^6$ yr. The last four panels show the temperature 
distribution, as in Figure 2, for the last calculated model. The size 
of the plots displays the whole computational grid: 100 pc $\times$ 1 kpc.
\label{fig3}}

\figcaption[gtt4.gif] 
{The same as Figure 3 for Z = 10Z$_{\odot}$. The evolutionary times of 
the first four panels is 
1.79 $\times$ 10$^5$ yr, 4.82 $\times$ 10$^5$ yr, 1.05 $\times$ 10$^6$ yr
and 1.39 $\times$ 10$^6$ yr, respectively.  
\label{fig4}}

\figcaption[gtt5.gif] 
{The same as Figure 3 for Z = 10Z$_{\odot}$. The calculation considers 
only two superstar clusters far from the symmetry axis. The
evolutionary times of the first four panels is 
1.8 $\times$ 10$^5$ yr, 4.94 $\times$ 10$^5$ yr, 1.05 $\times$ 10$^6$ yr
and 1.4 $\times$ 10$^6$ yr, respectively.  
\label{fig5}}

We thank an anonymous referee for multiple suggestions and comments.
The authors  acknowledge support from  CONACYT - M\'exico, research 
grant 36132-E and from the Spanish Consejo Superior de Investigaciones 
Cient\'\i{}ficas, grant AYA2001 - 3939.

\end{document}